\documentclass[pra,letterpaper,twocolumn,showpacs,superscriptaddress,floatfix]{revtex4}
\usepackage{graphicx,psfrag,amsmath,amssymb,amsfonts,bbm,latexsym,color,dcolumn,bm}

\begin{document} 

\title{Ehrenfest Dynamics and Frictionless Cooling Methods}
\author{Stephen Choi}

\affiliation{Department of Physics, University of Massachusetts, Boston, MA 02125, USA}

\author{Roberto Onofrio}

\affiliation{\mbox{Dipartimento di Fisica e Astronomia ``Galileo Galilei'', Universit\`a  di Padova, 
Via Marzolo 8, Padova 35131, Italy}}

\affiliation{ITAMP, Harvard-Smithsonian Center for Astrophysics, 60 Garden Street, Cambridge, MA 02138, USA}

\author{Bala Sundaram}

\affiliation{Department of Physics, University of Massachusetts, Boston, MA 02125, USA}

\begin{abstract}
Recently introduced methods which result in shortcuts to adiabaticity,
particularly in the context of frictionless cooling, are rederived and 
discussed in the framework of an approach based on Ehrenfest dynamics.  
This construction provides physical insights into the emergence of
the Ermakov equation, the choice of its boundary conditions, and the
use of minimum uncertainty states as indicators of the efficiency of the procedure. 
Additionally, it facilitates the extension of frictionless cooling to
more general situations of physical relevance, such as optical dipole trapping schemes. 
In this context, we discuss frictionless cooling in the
  short-time limit, a complementary case to the one considered in the
  literature, making explicit the limitations intrinsic to the
  technique when the full three-dimensional case is analyzed.
\end{abstract}


\maketitle

\section{Introduction}

Cooling quantum systems to the lowest reachable temperatures is a goal with 
motivations arising both from fundamental and practical considerations. 
The ultimate control of microscopic systems in the ultracold regime
allows for the full exploitation of quantum technologies as well as the understanding of the attainability of zero temperature. 
Much of the focus in cooling has been on the manipulation of populations in energy
levels, through coupling to external reservoirs, with the goal of increasing 
occupancy of the lowest energy levels. However, an alternative
strategy consists of temporally manipulating a parameter of
the system Hamiltonian to reduce the energy content of each level
while keeping occupancy invariant. These strategies are  ``adiabatic'' 
in the fullest sense, both thermodynamic and quantum, as they neither 
change the entropy nor the energy distribution of the system.

Usual adiabatic protocols require changing a Hamiltonian parameter
such that the rate of change of the parameter times its 
duration is much smaller than its absolute initial value. 
This requirement severely constrains the cooling power, making 
adiabatic cooling techniques less favorable in those contexts 
where, in addition to the intrinsic dynamics, sources of decoherence 
and noise exist, hindering desired tasks such as quantum computation or simulation.
Techniques to achieve shortcuts to adiabaticity relax this condition
by only requiring constancy of entropy at the initial and final times,
 but not necessarily at intermediate times. 
By identifying a global time invariant, temporal trajectories of the 
Hamiltonian parameter can be found such that the final population
distribution equals the initial one while all the energy eigenvalues 
are scaled down by a common factor.
 This cooling technique, known as ``frictionless cooling,'' sketched in the conclusions of 
Ref.~\cite{Yuce}, has been extended to an atomic framework~\cite{Chen} leading to an increasing number of applications 
~\cite{Schaffrev,Torronteguirev}. 
Adiabatic or frictionless cooling does not reduce the entropy of the 
system under consideration~\cite{Ketterle}, making it 
ineffective for situations in which entropy and phase space density play the 
leading role such as in triggering phase transitions. However, these techniques do reduce the  
temperature, with all the associated benefits in terms of state
preparation~\cite{Masudanote}. 
Examples include efficient fast decompression of ${}^{87}$Rb atoms in 
normal~\cite{Schaff0} and Bose-condensed~\cite{Schaff} states, which
have been experimentally demonstrated, and detailed proposals for efficient fast atomic transport~\cite{Torrontegui} 
and optimized sympathetic cooling~\cite{Choionba1}.

In harmonic potentials, frictionless cooling is achieved by choosing 
the variability of the trap frequency as specified by the solution of 
a second order differential equation, the Ermakov equation. 
In our earlier work~\cite{Choionba2}, we had addressed  
the robustness of this protocol to realistic sources of uncertainties 
and errors, and had shown that the Ermakov solution leads to minimum
uncertainty wavepackets at both initial and final times. Additionally,
this protocol resulted in squeezing of the momentum variance, 
formally parametrized through Bogoliubov transformations, during the dynamical evolution.
This allowed us to use the degree of squeezing seen in the evolved 
solution as an effective measure of fidelity. However, the reasons
behind the emergence of minimum uncertainty states at the final 
time, and not during intermediate times, were not made explicit. 
In this paper, we fill in some of these gaps in understanding this
technique by considering the Ehrenfest dynamics of the Heisenberg 
operator equations for the time-dependent harmonic oscillator \cite{Ehrenfest,ChemLit}.  
In particular, the physical interpretation of variables in the 
Ermakov construction, the choice of boundary conditions, and the role 
of squeezing in the solution all become manifest. 
We also extend the construction to a special case where the dynamics 
is restricted to the class of generalized Gaussian states, which
results in the so-called Effective Gaussian Dynamics (EGD) approach \cite{PS1,PS2}.  
While both Ehrenfest dynamics and EGD approach are exact for quadratic
potentials, more generally they are known to provide approximate 
but valid results specifically at short times \cite{PWMBS}, which
makes them well-suited for application to frictionless cooling methods
in their fastest regime. 

The paper is organized as follows. In Section II we provide a brief
introduction to the general Heisenberg operator approach and the special EGD case.
 In Section III this formalism is applied to the important case of
  harmonic oscillator or quadratic potential and provide a brief discussion of the resulting set of equations.
In Section IV we explicitly show the connection of our construction
to frictionless cooling in the case of harmonic potentials. 
In Section V we discuss possible extensions within the EGD framework, with 
particular emphasis on the experimentally relevant case of optical
dipole trapping. Finally we conclude with some qualitative insights on 
the usefulness of Heisenberg operator equations to further expand the concept 
of frictionless cooling.
  
\section{Heisenberg operator equations and Ehrenfest dynamics}

We begin our analysis considering a general time-dependent Hamiltonian 
$\hat{H} = \hat{p}^2/2m + V(\hat{x},t)$. The associated Heisenberg equations are:
\begin{eqnarray}
\frac{d\hat{x}}{dt} &=&  \frac{\hat{p}}{m} \;, \nonumber \\
\frac{d\hat{p}}{dt} &=& - \frac{\partial V(\hat{x},t)}{\partial x} \;.
\end{eqnarray}
Writing each operator $\hat{A} =  \langle \hat{A} \rangle +\Delta
\hat{A}$, where $\langle \ldots \rangle$ denotes the expectation value so that
$\langle \Delta \hat{A} \rangle =0$, one can Taylor expand the
potential $V(\hat{x},t)$ about $\langle \hat{x} \rangle$ resulting in 
the following pair of Ehrenfest equations 
\begin{eqnarray}
\frac{d \langle \hat{x} \rangle}{dt} &=& \frac{\langle \hat{p} \rangle}{m} \;,  \label{Ehrenfest1} \\
\frac{d \langle \hat{p} \rangle}{dt} &=& - \sum_{n=0}^{\infty}
\frac{1}{n!} V^{(n+1)} (\langle \hat{x} \rangle) \langle \Delta \hat{x}^n \rangle \;,
\end{eqnarray}
where $V^{(n)} = \partial^{n} V/\partial x^{n}$. Writing down the
corresponding evolution equations for $\langle \Delta\hat{x}^n\rangle$ 
leads to an infinite hierarchy of equations. 
It is worth noting here that the functional form of $V(\hat{x},t)$ is
important in coupling higher moments to the evolution of the centroid variables. 
 The infinite set of moment equations are typically
truncated using one of a number of possible approximations, largely determined by the nature of
the problem being addressed~\cite{Hanggi}. 

Of course, one can truncate the infinite equations order by order
which is equivalent to approximating any potential as a polynomial, 
where the degree is related to the order of the correlations that are
retained. However, this truncation leads to the coupling of moments
across different orders as soon as one goes beyond the second order.  
This results in higher-order moments becoming dynamically significant 
even if they were initially (at $t=0$) zero. 
The only exception to this behavior is the special case of a quadratic potential where, 
at each order, the moment equations depend only on other moments of the same order. 
This is readily illustrated by writing down the second-order contributions
\begin{eqnarray}
\frac{d \langle \Delta\hat{x}^2 \rangle}{dt} &=& \frac{1}{m} \langle
\Delta \hat{x} \Delta \hat{p} + \Delta\hat{p} \Delta\hat{x} 
\rangle,  \label{Ehrenfest2} \\
\frac{d \langle \Delta \hat{x} \Delta \hat{p} + \Delta\hat{p}
  \Delta\hat{x} \rangle}{dt} &=& \frac{2}{m} \langle \Delta\hat{p}^2 
\rangle -2V^{(2)}(\langle \hat{x} \rangle) \langle \Delta \hat{x}^2
\rangle, \label{Ehrenfest3} \\
\frac{d \langle \Delta\hat{p}^2 \rangle}{dt} &=& -V^{(2)}(\langle
\hat{x} \rangle) \langle \Delta \hat{x} \Delta \hat{p} + 
\Delta\hat{p} \Delta\hat{x} \rangle. 
\label{Ehrenfest4} 
\end{eqnarray}
The more general result for the evolution equations for higher-order
moments can be readily written down, although with tedium increasing
progressively with each order. 

It is clear that straight truncation up to second order may not be
effective for arbitrary, non-quadratic potentials. Improved accuracy 
in arbitrary potentials while keeping the number of Ehrenfest equations finite is desirable. 
In this regard, another related method to truncate the infinite
hierarchy of moment equations involves the use of a time-dependent 
variational approach in which the state of the system is assumed to 
remain in a general Gaussian form. 
The major implication
of the Gaussian approximation is that higher-order correlations can
be expressed in terms of one and two-point correlations alone, leading 
to a dramatic truncation in the space of variables. 
Also, given that 
arbitrary operators $A$, $B$, and $C$ with $[A,B] = i C$ implies \cite{Uncertainty}
\begin{equation}
\langle A^2 \rangle \langle B^2 \rangle \geq \frac{1}{4}\langle C \rangle^2
+ \frac{1}{4} \langle AB + BA \rangle^2  \;,
\end{equation}
the general Gaussian form obeys the uncertainty relation
\begin{equation}
\langle \Delta \hat{x}^2 \rangle \langle \Delta \hat{p}^2 \rangle = \frac{\hbar^2}{4}  +
\frac{1}{4} \langle \Delta \hat{x} \Delta \hat{p} + 
\Delta \hat{p} \Delta \hat{x} \rangle^2\;,  \label{HUP}
\end{equation}
which helps to simplify the Ehrenfest equations. The resulting EGD  is represented by \cite{PS1,PS2}
\begin{eqnarray}
\frac{d  x }{dt}  & = &  \frac{ p}{m}  \;, \label{dx0}  \\
\frac{d p  }{dt}  &  = & - \sum_{n = 0}^{\infty}
 V^{(2n+1)}( x ) \frac{\rho^{2n}}{n! 2^{n}} \;, \label{dp0} \\
\frac{d \rho}{dt}  & = & \frac{\Pi}{m}  \;,  \label{drho0} \\
\frac{d \Pi}{dt}  & = & \frac{\hbar^2}{4 m \rho^3} -  
\sum_{n =0}^{\infty}  V^{(2n+2)}( x
) \frac{\rho^{2n+1}}{n! 2^n} \;, \label{dPi0}
\end{eqnarray}
where $x \equiv \langle \hat{x} \rangle$ and $p \equiv \langle \hat{p}
\rangle$ are the expectation values of position and momentum, respectively. 
Here, odd cumulants are identically zero and even cumulants can be
written in terms of variable $\rho$ as $\langle \Delta \hat{x}^{2n} \rangle= 
\rho^{2n}2n!/(2^{n}n!)$. We also introduce a new variable 
$\Pi = \langle \Delta \hat{x} \Delta \hat{p} + \Delta \hat{p} \Delta
\hat{x} \rangle/2 \rho$ 
which, as is clear from its definition, reflects the correlation between
$\Delta \hat{x}$ and $\Delta \hat{p}$. 
Together, the four equations of motion fully describe the evolution of
both the centroid and the spreading of the wave packet. 

Before considering the case of the time-dependent harmonic oscillator which serves as a useful
paradigm for frictionless cooling methods, an important difference between EGD
and the second-order truncation methods is worth noting. The second-order truncation 
consists of locally approximating an arbitrary potential by an effective quadratic one and, in keeping
with the Heisenberg picture, places no restrictions on the wavefunction. By contrast, the EGD method
assumes a Gaussian state which in terms of the potential results in a polynomial approximation 
involving only even powers. Thus, in the general case, these two approximations are different and are
valid for differing evolution times. As we see in the next section, an exception is the case of a harmonic potential 
where the two methods converge. 

\section{Ehrenfest dynamics for a harmonic oscillator}

Given our motivation of connecting the Ehrenfest equations to those
seen in frictionless cooling methods, we specifically consider the
case of a harmonic trap with a time-dependent angular frequency,  
{\it i.e.} $V(x,t)  =  \frac{1}{2} m\omega^2(t) x^2$. 
In this instance, both the more general approach Eqs. 
(\ref{Ehrenfest1})-(\ref{Ehrenfest4}) and the EGD 
yield identical and considerably simplified relations
\begin{eqnarray}
\frac{d x  }{dt} & = & \frac{ p }{m}  \;,  \label{dx} \\
\frac{d  p }{dt} &  = &  - m \omega^2(t)  x  \;, \label{dp1} \\
 \frac{d \rho}{dt} & = &  \frac{\Pi}{m}  \;,  \label{drho}  \\
\frac{d \Pi}{dt}   & = &  \frac{\hbar^2}{4m  \rho^3}   - m \omega^2(t)  \rho \;. \label{dPi1} 
\end{eqnarray}
It is clear that in this specific case of the harmonic oscillator, Eqs. (\ref{dx}) and (\ref{dp1}), 
and Eqs. (\ref{drho}) and (\ref{dPi1}) are completely decoupled, and one can look at the evolution 
of the mean position and momentum completely independently of the evolution of the respective 
fluctuations, allowing for an easy numerical integration. 
Also, it is interesting that, due to the structure of the equations,
Eq. (\ref{dPi1}) can describe the case of an external potential that includes 
up to a linear term in position via an arbitrary, in general time-dependent, 
constant $\beta(t)$, {\it i.e.} $V(x,t)  =  \frac{1}{2} m \omega^2(t) x^2  + \beta(t) x$. 

The decoupling of the centroid and fluctuation relations allows the 
recasting of the problem in terms of a higher-dimensional system where 
the fluctuation and average variables are treated on an equal footing. 
In terms of the canonical variable pairs $(x,p)$ and $(\rho,\Pi)$, the
extended Hamiltonian $H_{ext} \equiv \langle H \rangle$ is given by
$H_{ext} = H_{ px } + H_{\Pi \rho}$ where
\begin{eqnarray}
H_{ px } & =  & \frac{p^2}{2m} + \frac{1}{2}
m\omega^2(t) x^2  \;,   \\
H_{ \Pi \rho} & =  &  \frac{\Pi^2}{2m}
+ \frac{\hbar^2}{8m \rho^2}  + \frac{1}{2}m\omega^2(t) \rho^2  \;.
\label{Hrho}
\end{eqnarray}

A number of remarks are now in order. First, the extended
Hamiltonian has a centrifugal barrier which prevents $\rho  \rightarrow 0$ 
except in the trivial, classical limit in which we can assume $\hbar \rightarrow 0$. 
Second, the quantum correction preventing null position fluctuations
is proportional to the second power of the Planck constant,
consistently with the osmotic term present in the Madelung-Bohm form of the
Schr\"odinger equation ~\cite{Madelung}. 
Third, it is simple to show that the effective potential for the
fluctuating part -- if expanded around the minimum for an oscillator
with constant frequency -- generates small fluctuations with 
average value oscillating harmonically in time (`breathing' modes)
with frequency twice the oscillator frequency. This fact is exploited
in stroboscopic quantum nondemolition measurements of position ~\cite{Braginsky,Thorne}. 
Finally, the decoupling of the first and second moments is a 
particular case, for a single particle, of Kohn's theorem valid in the 
more general situation of an interacting many-body system ~\cite{Kohn}.

\section{Connection to frictionless cooling}

The goal of this Section is to derive from the Ehrenfest perspective the results of 
frictionless cooling as described in \cite{Chen}. 
In order to make this contribution self-consistent, we briefly recall that the 
Ermakov trajectory, prescribing the time variation of the harmonic
trapping frequency necessary for frictionless cooling, arises from the Lewis-Riesenfeld invariant 
$I(t)=1/2 [{\pi^2}/{m}+m\omega_0^2 (q/b)^2]$,  
where $\pi  = b p - m \dot{b} q$.  The invariant is obtained, as first shown by Ermakov 
\cite{Ermakov}, by introducing an auxiliary equation to the Newtonian equation for the 
harmonic oscillator, which determines the evolution of a scaling factor $b$ 
related to the position variable $x = q/b$. 
The resulting Ermakov equation is given by
\begin{equation}
\ddot{b} + \omega^2(t) b = \omega_0^2/b^3\;.
\end{equation}

\begin{figure}[t]
 \begin{center}
\centerline{\includegraphics[height=7cm]{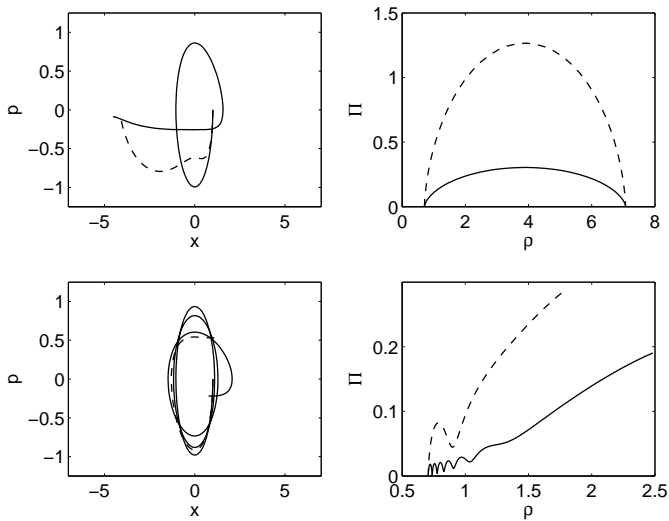}} 
\caption{Generalized phase space plots of the Ehrenfest dynamics in the 
case of the harmonic oscillator.  Here and in all the subsequent
  figures, all variables are given in harmonic oscillator units, with 
$\hbar = m =1$. By definition $\Pi$ has units of harmonic oscillator
  momentum.   In the top row we depict the case of
a harmonic oscillator driven through a Ermakov trajectory, with the
solid line corresponding to $t_f=$25 ms, and the dashed line to a
faster frictionless cooling occurring in $t_f=$6 ms (see also Fig. 4 
in \cite{Ibanez}). 
On the left we report the centroid dynamics, $p$ {\it vs.} $x$, on 
the right column the fluctuation dynamics, $\Pi$ {\it vs.} $\rho$.  
For comparison, in the bottom row we report the same quantities for 
the case of a linear ramping-down of the frequency occurring in the 
same time durations, showing that in the 25 ms case the harmonic
oscillator performs several cycles with respect to the corresponding Ermakov trajectory.
Notice that the Ermakov trajectories in the fluctuational phase space
always lead to a final minimum uncertainty state $\Pi=0$, unlike the linear
ramp-down trajectories.} 
\end{center}
\vspace{-0.75cm}
\end{figure}

\begin{figure}[t]
\begin{center}
\centerline{\includegraphics[height=7cm]{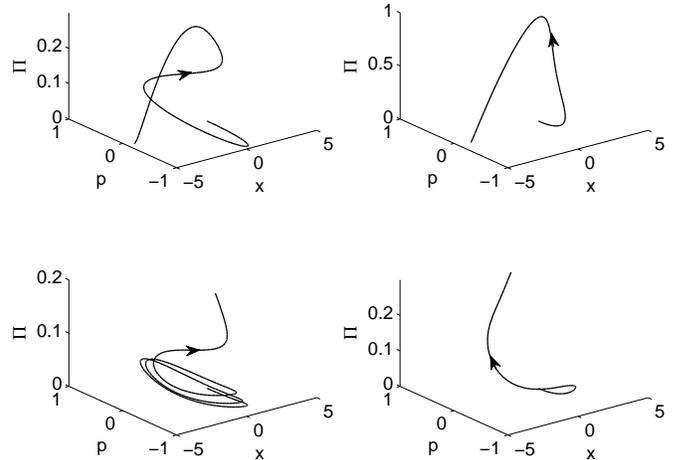}} 
\caption{Plot of $\Pi$ {\it vs.} $p$ and $x$ for the cases shown in 
Fig. 1.  The left column corresponds for $t_f=$25 ms, the right column 
to $t_f=$6 ms, the top row is for the Ermakov trajectory, the 
bottom row for the corresponding linear ramping with the same $t_f$. 
An increasing $\Pi$ as seen in the linear ramping cases leads to 
larger squeezing of the final wavefunction and an increasing departure
from the desired minimum uncertainty state.}
\end{center}
\vspace{-0.75cm}
\end{figure}

In order for the invariant $I(t)$ to commute with the Hamiltonian (given as Eq. (3) in \cite{Chen})
it is necessary that at $t = 0$,  $b(0) = 1$ and $\dot{b}(0) = 0$. 
Also  the choice $\ddot{b}(0) = 0$ ensures that $\omega(0)  = \omega_0$.  
At $t = t_f$ the conditions $b(t_f) = (\omega_0/\omega_f)^{1/2}$,
$\dot{b}(t_f) = 0$, and $\ddot{b}(t_f) = 0$ are imposed.
These make sure that $I(t)$ commutes with the Hamiltonian at $t=t_f$
and that $\omega(t_f) = \omega_f$. 
These six boundary conditions suggests a fifth-order polynomial Ansatz for $b(t)$, 
facilitating a solution of the Ermakov equation for $b(t)$, and
subsequently the explicit Ermakov trajectory for the angular frequency
$\omega(t)$.

The scaling factor $b(t)$ is proportional to the standard deviation of
the wave function $\sigma(t)$, such that for the ground state $n = 0$, 
$b(t) = \sigma(t)/(\hbar/2 m \omega_0)^{1/2}$, {\it  i.e.} the
standard deviation expressed in harmonic oscillator length units, 
$a_0=(\hbar/2m\omega_0)^{1/2}$ \cite{Chen}. 
This allows us to make the identification with the parameter $\rho$ 
in  Eqs. (\ref{drho}-\ref{dPi1}) $a_0 b(t) \equiv  \rho(t) \equiv \sigma(t)$. 
In particular, one can condense the four equations  Eqs. (\ref{dx}-\ref{dp1}), 
and Eqs. (\ref{drho}-\ref{dPi1}) into two second-order differential equations
\begin{eqnarray}
\frac{d^2   x }{dt^2} &  = &  - \omega^2(t) x  \;,  \label{dp}\\
\frac{d^2 \rho}{dt^2}   & = &  \frac{\hbar^2}{4 m^2 \rho^3 } -  \omega^2(t)  \rho \;, \label{dPi}
\end{eqnarray}
where, again, Eq. (\ref{dp}) is completely decoupled from Eq. (\ref{dPi}). 
The first is nothing but the Newton equation for the harmonic oscillator while
the equation for the spreading of the wave packet, Eq. (\ref{dPi}), can be rearranged as 
\begin{equation}
\ddot{\rho}+ \omega^2(t)\rho=(\hbar^2/4m^2)/\rho^3,
\end{equation}
which  coincides with the Ermakov equation by identifying $\omega_0^2=
\hbar^2/(4m^2 a_0^4)$ along with $b = \rho/a_0$.
Therefore the Ermakov equation obtained through a proper, but less
physically insightful, identification of an invariant naturally emerges here 
from the Ehrenfest formulation\cite{Rajagopal}. The identification of $b$ with $\rho$
also means that is not possible to impose Dirichlet boundary conditions, 
since $\rho$ cannot be zero. Solving the equations under Neumann boundary conditions
makes clear the underpinning for the minimum uncertainty state seen at $t = t_f$.

Additionally, one can identify $m a_0 \dot{b}(t) = \Pi(t)$.  
The  boundary conditions imposed $\dot{b}(0) =\dot{b}(t_f) = 0$
translate into $\Pi(0) = \Pi(t_f) = 0$. 
In our earlier work \cite{Choionba2}, we found that the Ermakov
trajectory requires a minimum uncertainty wavepacket at both initial and final times 
$t = 0$ and $t = t_f$ which, in the terminology of the moments, equates to $\Pi = 0$ at these times. 
So when we follow the Ermakov trajectory $\omega(t)$ it is now not
surprising that a minimum uncertainty state is achieved at $t = t_f$, 
as this is {\sl enforced} by the {\sl unique} choice of boundary conditions. 
Notice also that the momentum variance, related to the temperature of the ultracold gas,
can be written in terms of $\rho$ and $\Pi$ as
$\langle \Delta \hat{p}^{2} \rangle  = \hbar^2/(4\rho^2)+\Pi^2$,
from the uncertainty relation Eq. (\ref{HUP}), which shows that
momentum fluctuations are reduced if $\rho$ is made large  and, simultaneously, $\Pi=0$. 
This relationship makes it clear that a mere increase of the position
variance $\rho$ ({\it e.g.} by directly relaxing the trap frequency)
with the goal of reducing the corresponding momentum variance does not
necessarily work unless the system reaches a minimum uncertainty
state at the final time. Removal of squeezing correlations through $\Pi = 0$ is therefore the 
key step in a frictionless cooling scheme.

Furthermore, the boundary conditions  $\ddot{b}(0) = \ddot{b}(t_f) = 0$ imply
$\dot{\Pi}(0) = \dot{\Pi}(t_f) = 0$, and Eq. (\ref{dPi}) then gives
the width $\rho$ consistent with the eigenstate of a harmonic trap
with the correct angular frequencies at $t=0$ and $t=t_f$, as it should. 
It is worth noting that in the true adiabatic limit one starts with a
minimum uncertainty wavepacket (associated with the initial frequency
$\omega(0)$) and it remains a minimum uncertainty packet for all times
during the evolution. This requires $\Pi=0$ and $d\Pi/dt=0$ for all
$t$. Any deviation from this leads to diabatic transitions which can be countered by
non-zero values of $\Pi$.  It is interesting that this is precisely  
the counter-diabatic anticommutator term  found in other approaches to
achieve shortcuts to adiabaticity \cite{Rice,CJ,DelCampo}.  

We have explicitly confirmed the validity of the Ehrenfest dynamics and EGD  for the time-dependent quadratic 
potential by using the numerically obtained Ermakov trajectory as the input for $\omega(t)$ in 
 Eqs (\ref{dx} - \ref{dPi1}). The time evolution of the variables was found to be identical 
to those obtained by  numerically integrating the full Sch\"{o}dinger Equation.  
This agreement holds even for trajectories involving short $t_f$ which includes 
an antitrapping stage. The Gaussian Ansatz of EGD implies that the wave function remains a coherent state even in 
the presence of an inverted trap as the time is too short for the wave 
function to start breaking up.

The phase space diagrams for $x$ {\it vs.} $p$ and $\rho$ {\it vs.}
$\Pi$ are shown in Fig. 1 for different cooling protocols. As in our previous work 
\cite{Choionba1,Choionba2}, we consider two representative final times in all of 
our simulations -- $t_f = 25$ ms and $t_f = 6$ ms -- where the 6 ms case involves an 
anti-trapping stage while the 25 ms case does not. 
We compare the Ermakov trajectory (top row) to the case of a quasi-adiabatic 
protocol obtained with a linear ramp of the frequency in the same time interval (bottom row).
As expected, the Ermakov trajectory gives $\Pi = 0$  at the end of the run, implying 
that the state returns to the minimum uncertainty state as a consequence of the imposed boundary 
conditions, while the linear ramp case shows non-zero final $\Pi$.  
A three-dimensional plot is presented in Fig. 2 to show more explicitly the
evolution of squeezing along the trajectories. A more direct
representation of this dynamics in terms of the corresponding Wigner function has been 
discussed in Ref.~ \cite{Schuch}.

\section{Extensions to optical dipole trapping using Gaussian beams}

Fast expansion methods have also been discussed in the more realistic case
of optical Gaussian-beam traps in Ref.~\cite{Torrontegui1}. 
This is a more intriguing case than usual magnetic traps since,
  unlike the latter, trapping frequencies along the radial and axial 
direction (in the almost universally adopted confinement geometry with 
radial symmetry) cannot be independently controlled.
Nevertheless, optical dipole traps enjoy several advantages in ultracold atom
experiments, among these the possibility to trap spinor condensates,
the flexibility in independently using magnetic fields (for instance
to exploit magnetic Feshbach resonances), the possibility to trap atoms 
with no permament magnetic moment, and the possibility for atomic  
control with higher spatial and temporal resolution. It is therefore
worthwhile to discuss to what extent frictionless cooling techniques 
may be applied to this important class of trapping schemes.   
We will deal with the simplest situation of a single Gaussian beam of 
wavelength $\lambda$ red-detuned with respect to the dominant
atomic transition $\lambda_{\mathrm{at}}$. 
Taking into account the intensity profile of a Gaussian laser beam in the paraxial 
approximation, and identifying a symmetry axis in the direction of the
light propagation along the $z$ axis, the effective potential energy
felt by the atoms is given, in terms of the radial and axial coordinates $r$ and $z$, as
\begin{equation}
V_{\rm opt}(r,z,t) = V_0(t) \left [ 1 -  \frac{w^2_0}{ w^2(z)} e^{-2r^2/w^2(z)} \right ]  \;,
\label{OptPot}
\end{equation}
where $V_0(t) = 3 I_0(t)\lambda^3/(16 \pi^2 c \tau \delta)$, with
$I_0(t)$ the instantaneous beam intensity, $\tau$ the lifetime of the 
excited state, $\delta=\lambda-\lambda_{\mathrm{at}}$ the detuning between the 
light wavelength and the atomic transition wavelength, 
$w_0$ the beam waist, and $w(z) = w_0 \sqrt{1 + \left ( z/z_R \right )^2 }$ 
the spot size at location $z$, where $z_R = \pi w_0^2/\lambda$ is the Rayleigh range.  

On applying the EGD to this effectively 2D situation, one gets two sets of equations 
Eqs. (\ref{dx0} - \ref{dPi0}) for the radial and axial directions. 
In each direction, the expectation values and two-point correlations decouple and 
each gives rise to its own Ermakov-type fluctuation  equations. 
Explicit EGD equations  are given by evaluating Eqs. (\ref{dp0}) and
(\ref{dPi0}) using the Gaussian optical potential  of Eq. (\ref{OptPot}) 
for the experimentally reasonable case of $\langle r \rangle  = \langle z
\rangle  = 0$. From now on, we shall use the simplified notation $r \equiv \langle r
\rangle$ and $z \equiv \langle z \rangle$.

In the radial direction, the infinite series can be evaluated  using the
properties of Hermite polynomials to give $V_{opt}^{(2n+1)}(r)=0$
for all $n$. By using various identities involving the generalized 
Laguerre polynomials the series that includes the partial 
derivatives $V_{opt}^{(2n+2)}(r)$ can, after some work, be evaluated to finally yield
\begin{eqnarray}
\frac{d^2   r }{dt^2}  & = &  0 , \label{Newton_r} \\
\frac{d^2 \rho_r}{dt^2}   & = &   \frac{\hbar^2}{4 m^2 \rho_r^3}  -
\frac{4V_0 (t) }{m w_0^2}  
\rho_r \left [1 + \left ( \frac{2\rho_r}{w_0} \right )^{2} \right
]^{-3/2} . 
\label{ddrho_r} 
\end{eqnarray}
In the axial direction,  we have  instead a Lorentzian function in $z$
as our  $V_{opt}(r,z,t)$ and it turns out that, again,
$V_{opt}^{(2n+1)}(z)=0$ for all $n$,  while the partial
derivatives $V_{opt}^{(2n+2)}( z )$ are found to simplify  to give
\begin{eqnarray}
\frac{d^2   z }{dt^2} & = & 0 , \label{Newton_z} \\
\frac{d^2 \rho_z}{dt^2}   & = &   \frac{\hbar^2}{4 m^2 \rho_z^3}  -
\frac{V_0 (t) }{m z_R^2} 
\rho_z \sum_{n = 0}^{\infty} \frac{(2n +2)!}{n!} \left
(-\frac{\rho_z^2}{2 z_R^2} \right )^{n} .  
\label{ddrho_z}
\end{eqnarray}

\begin{figure}[t]
\begin{center}
\centerline{\includegraphics[height=7cm]{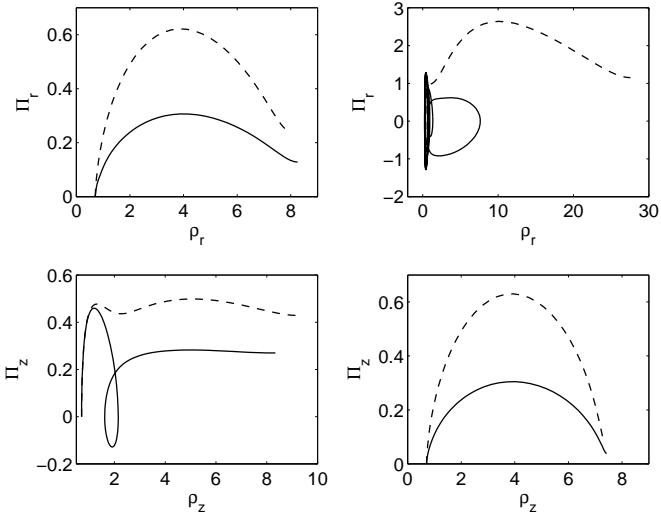}} 
\caption{Ehrenfest dynamics for the optical dipole potential 
as in Eq. (\ref{OptPot}) expressed through the fluctuative components
in the $\Pi$ {\it vs.} $\rho$ phase diagram. The top row represents 
the radial variables $\Pi_r$ and $\rho_r$, the bottom row is for the 
axial variables $\Pi_z$ and $\rho_z$. In the left column the 
Ermakov driving $\omega_r^2(t)$ is imposed on the radial frequency, in the 
right column the axial frequency is instead Ermakov-driven with $\omega_z^2(t)$.  
We have chosen a beam waist of $w_0=20a_0$, and as usual 
we consider two different times $t_f$=25 ms (solid line) 
and $t_f$=6 ms (dashed line).}  
\label{PiRho}
\end{center}
\vspace{-0.75cm}
\end{figure}

The Newtonian relations (\ref{Newton_r}) and (\ref{Newton_z}) make it clear that the expansion 
is around the equilibrium point $r=z=0$.
The corresponding extended Hamiltonians in the $\Pi$--$\rho$ space are given by 
\begin{eqnarray}
H_{\Pi_r \rho_r} & =  &  \frac{\Pi_r^2}{2m} +\frac{\hbar^2}{8 m
  \rho_r^2}  -  V_0(t)
\left [ 1+ \left ( \frac{2 \rho_r}{w_0} \right ) ^2 \right ]^{-1/2} ,  \label{Hrho_r} \\
H_{\Pi_z \rho_z} & =  &  \frac{\Pi_z^2}{2m} + \frac{\hbar^2}{8m
  \rho_z^2}  +   
\frac{V_0 (t) }{z_R^2}  \sum_{n = 0}^{\infty} \frac{(2n +1)!}{n!}  \nonumber \\
 & & \times  \left ( -\frac{1}{2 z_R^2} \right )^{n} \rho_z^{2n+2}  .  \label{Hrho_z} 
\end{eqnarray}

In the limit of small $\rho_r$ and $\rho_z$ (or large $w_0$ and $z_R$)
one can approximate Eqs. (\ref{ddrho_r}) and (\ref{ddrho_z}) to rederive the Ermakov equations
\begin{eqnarray}
\frac{d^2 \rho_r}{dt^2}   & = &   \frac{\hbar^2}{4 m^2 \rho_r^3}  - \frac{4V_0 (t)  }{m w_0^2} \rho_r  \;,\label{r_ermakov}  \\
\frac{d^2 \rho_z}{dt^2}   & = &   \frac{\hbar^2}{4 m^2 \rho_z^3}  -  \frac{2V_0(t)}{m z_R^2}  \rho_z   \;.\label{z_ermakov}
\end{eqnarray}
Comparing with the general form for the Ermakov Equation, we get the equivalent 
Ermakov angular frequencies $\omega_r^2(t) = 4V_0(t)/m w_0^2$ and $\omega_z^2(t) = 2V_0(t)/m z_R^2$. 
These are exactly the forms obtained by Muga and collaborators \cite{Torrontegui1} and 
despite the decoupling of the radial and axial coordinates, $\omega_r^2(t)$  and  $\omega_z^2(t)$ are 
related through $V_0(t)$
\begin{equation}
\omega_z^2(t) = \frac{w_0^2}{2z_R^2} \omega_r^2(t) \;.  
\label{condition}
\end{equation}

\begin{figure}[t]
\begin{center}
\centerline{\includegraphics[height=7cm]{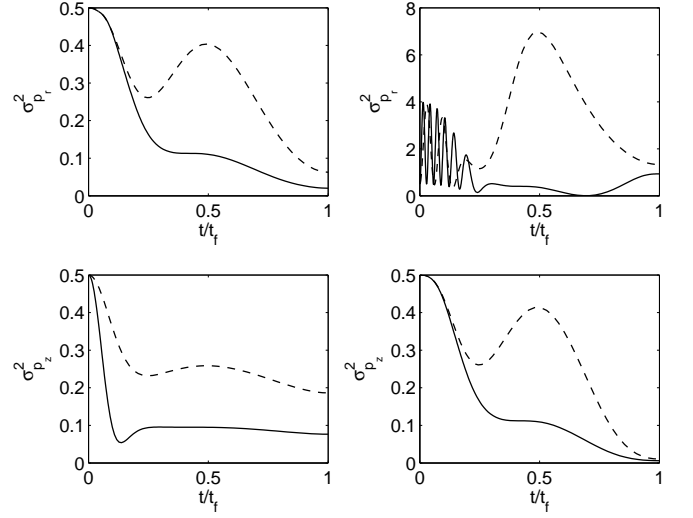}} 
\caption{Time evolution of the momentum variance for the same 
arrangement as in Fig. \ref{PiRho}}. 
\label{temp}
\end{center}
\vspace{-0.75cm}
\end{figure}

This implies that optimal fast-adiabatic cooling can only be achieved
in either one of the radial and axial directions for a given run
{\it i.e.} the other direction may not be simultaneously cooled
optimally, hopefully undergoing a sort of indirect, passive frictionless cooling. 

\begin{table*}[t]
\centering 
\begin{tabular}{c | c | c c c c | c c c c} 
\hline\hline 
 &    &      & Master &  Radial  &   &   &  Master  & Axial & \\
$w_0$ & $t_f$(ms) &  $\langle \Delta \hat{p}_r^2 \rangle$ & $\Pi_r$ &
 $\langle \Delta \hat{p}_z^2 \rangle$ & $\Pi_z$  &  $\langle \Delta
 \hat{p}_r^2 \rangle$ & $\Pi_r$ &   $\langle \Delta \hat{p}_z^2
 \rangle$ & $\Pi_z$   \\ 
[1.0ex] 
\hline
$20$ & $25$ & 0.0202  & 0.1285  &  0.0764  & 0.2699 & 0.9337 &  -0.4687  &  0.0060 &  0.0386 \\ 
$20$ & $6$  & 0.0627  & 0.2425  & 0.1864 & 0.4283 & 1.3265  & 1.1516 &  0.0104 &  0.0759  \\
$200$ & $25$ &  0.0050 & 0.0017 &  0.0769  &  0.2711 &  1.0618&    1.0268&   0.0050 &   0.0004 \\ 
$200$ & $6$  &   0.0050 & 0.0035  &   0.1733 &  0.4128 &   34.5122  &
 5.8745  & 0.0050   & 0.0009 \\ 
[1.0ex] 
\hline\hline 
\end{tabular}
\caption{Final radial and axial $\langle \Delta \hat{p}^2 \rangle$ and
$\Pi$ for the {\sl master radial} and {\sl master axial} cases.  
 All variables are given in harmonic oscillator units, with $\hbar = m =1$. 
In particular, the beam waist $w_0$ is expressed in units of the harmonic oscillator 
length $a_0=(\hbar/2m\omega_0)^{1/2}$ which, with $\omega_0/2\pi=$ 250 Hz, corresponds to
$a_0=0.95~\mu$m and $0.49~\mu$m in the most representative examples of ${}^{23}$Na and ${}^{87}$Rb atoms, respectively.}
\label{table1}
\end{table*}

We numerically simulate the full EGD relations, Eqs.  (\ref{ddrho_r})
and (\ref{ddrho_z}), and consider two cases, which are motivated by the relationship
between axial and radial trap frequencies. 
The first is the {\sl master radial} case when $\omega_r^2(t)$ is set to
be the Ermakov trajectory (and hence $\omega_z^2(t)$ follows a modified trajectory as given in
Eq. (\ref{condition})), and the second is the {\sl master axial} case for the inverse situation.
For illustrative purposes, we shall consider here the less ideal case
of small $w_0$ and $z_R$, {\it i.e.} away from the Ermakov limits of Eqs. (\ref{r_ermakov})
and (\ref{z_ermakov}) in our numerical simulations.  
We shall take  the minimum Rayleigh length that permits paraxial
approximation, $z_R=2 w_0$ (typically $z_R$ is much larger, for example 
$z_R=24w_0$ in \cite{Torrontegui1}). Also we choose the small beam waist of
$w_0$=20 and 200 harmonic oscillator lengths, based on the fact that the size of
a typical beam waist can range from $w_0=8 \mu$m (as in \cite{Torrontegui1}
which is of the order 53 harmonic oscillator lengths for ${}^{87}$Rb atoms) to 
sizes of order 0.3 mm (corresponding to roughly $2,000$ harmonic oscillator lengths).  

Figure \ref{PiRho} shows the $\Pi$ {\it vs.} $\rho$ phase diagram,
where the top row presents the phase space diagram  for the radial variables
$\Pi_r$ and $\rho_r$ and the bottom row is for the axial variables
$\Pi_z$ and $\rho_z$.  In the left column the {\sl master radial} case
is depicted, while the right column is describing the {\sl master axial} case.
Fig. \ref{temp} has the same arrangement as Fig. \ref{PiRho} except
that the momentum variance time evolution is presented.  
The results shown in Figs. \ref{PiRho} and \ref{temp} are obtained 
for a beam waist $w_0 = 20 a_0$. A tenfold increase in the beam waist  
results in the same qualitative behavior and the corresponding values 
for $\langle \Delta \hat{p}^2 \rangle$ and $\Pi$ are presented in Table \ref{table1}.

From the results we can see better {\sl master axial} performance both
in terms of the reduction in the momentum variance and the restoration
of the minimum uncertainty state indicated by $\Pi$ approaching zero. 
The results are practically indistinguishable from simulating the
Ermakov limit, Eq. (\ref{z_ermakov}). This is to be expected, since,
even with the deliberately less ideal choice of $w_0=20 a_0$ and $z_R = 2w_0$,  
the magnitudes of the coefficient of $\rho_z^{2n}$ in the series of 
Eq. (\ref{ddrho_z}) quickly drops to zero for increasing $n$. For 
instance, up to $n=5$ the magnitudes are: $2$, $7.5 \times 10^{-3}$,  
$3.52 \times 10^{-5}$, $2.05 \times 10^{-7}$ , $1.44 \times 10^{-9}$.   
On the other hand, for efficient cooling in both directions, the 
{\sl master radial} case with $t_f$=25 ms works better since the 
performance in the axial direction is not as severely compromised 
as in the inverse case. In general the $t_f$=25 ms cases yields 
better results in terms of cooling, being closer to $\Pi=0$.

We notice that in the top right hand panel of Fig. ~\ref{PiRho} 
($\Pi_r$ vs. $\rho_r$ for the {\sl master axial} arrangement) the 
phase space trajectory for $t_f$=6 ms covers a relatively large area
compared to the corresponding counterpart in the bottom left panel
($\Pi_z$ vs. $\rho_z$ for the {\sl master radial} arrangement). 
This can be understood from the difference in the extended Hamiltonians
Eq. (\ref{Hrho_r}) and (\ref{Hrho_z}). 
If plotted, the potential for the axial Hamiltonian overlaps very
closely to the case of a quadratic potential, while the potential for
the radial Hamiltonian deviates from the quadratic case as $\rho_r$
becomes larger, such that the potential decreases with a greater
(negative) slope compared to the quadratic case. This makes it easier
for the $\rho_r$ variable to extend to a greater distance from the
origin, especially with an antitrapping stage included in the $t_f$=6
ms case.

Another observation is the attainment of negative $\Pi$ in
Fig. \ref{PiRho} for $t_f$=25 ms in the ``subordinate'' cases (axial
under {\sl master radial} situation and vice versa).  One can understand
the negative $\Pi$ from the extended Hamiltonian as a function of
$\Pi$ and $\rho$, where, in the $\Pi$ direction the function is simply
parabolic and not multiplied by the Ermakov trajectory $\omega(t)$.  
Initially $\Pi=0$ then it grows over time due to squeezing  $\Pi>0$,  
{\it i.e.} rolls ``uphill'' and then it turns around to roll back down. 
For the longer non-optimal time of $t_f$=25 ms, it seems reasonable
that there is enough time to roll up to the other side of the hill
attaining $\Pi<0$. Physically, from the evolution equations,
negative $\Pi$ is seen to result in a decrease in the corresponding
$\rho$ variable. This means a reduction in the spatial width or the
``contractive" behavior associated with the so-called twisted coherent
states introduced in \cite{Yuen} (see also \cite{LorRob} for 
the contractive behavior of Schr\"{o}dinger cat states). 
The negative values of $\Pi$ do not impact the uncertainty relation 
since the latter depends on $\Pi^2$.

\section{Conclusions}

We have discussed  frictionless cooling in terms of the Ehrenfest
dynamics, getting more physical insight into the detailed nature of
the cooling process, and analyzed the relevant example of an optical 
dipole trap in the short-time duration regime which is complementary to 
the analysis reported in \cite{Torrontegui1}. This approach is also 
a simpler alternative to the search for Ermakov invariants in higher 
dimensional spaces as discussed in \cite{Leach} since in our framework 
the time-dependent frequencies are related to second derivatives of 
the potential evaluated at the expectation values.  

The fact that the Ermakov equation emerges for the case of a harmonic potential
via the application of EGD in which a Gaussian
wave packet Ansatz is imposed throughout the evolution is consistent with the
concept of adiabatic following, where an energy eigenstate remains so 
throughout the evolution. One may therefore view  the EGD as
generating a subspace of solutions that ``simulate''  the behavior of
adiabatic following. Alternatively,  it can be viewed as a way to 
naturally include the counter-diabatic term in the Hamiltonian \cite{Rice,CJ,DelCampo}. 
Formally, one should be able to generalize this idea to  any arbitrary 
eigenstates and trapping potentials. 
As mentioned above, the Gaussian wave packet in a quadratic potential
is the only case that involves a small number of tractable, closed set of equations
as an infinite chain of equations involving all cumulants results in
other situations (for instance see  \cite{Hasegawa} for the 
application to a double-well system). However, it is reassuring that 
higher order cumulants do not play a significant role for short times, and therefore 
should not make invalid the dynamics in situations in which a very short duration of the protocol 
is chosen. In this sense, and noting that solutions which are less than absolutely optimal 
may be sufficient for some cooling situations, we suggest that this ``Ansatz-enforced'' shortcut 
to adiabaticity may be applicable in more general situations. 
The procedure would start by truncating the Taylor expansion around the centroid in the
Ehrenfest equations to $N$ terms, with the truncation being exact for
potentials described by an $N$th order polynomial.  This would result
in a finite number of moment equations to be satisfied,  instead of a
single Ermakov Equation as in the case of a Gaussian Ansatz in a
harmonic potential. One can proceed by solving iteratively a truncated set of cumulant equations with an appropriate Ansatz imposed, using 
progressively more equations for higher accuracy. The high order moments could additionally be
controlled by the imposition of appropriate boundary conditions. 

Finally, it is noted that having $\omega(0)  \ll  \omega(t_f)$ one should also obtain
a ``fast-adiabatic heating'' effect, and this observation is more
transparent in the Ehrenfest framework as this corresponds to generalized 
phase space trajectories for the cumulants related to fast-adiabatic cooling via time-reversal. 
The added ability to deal with more general potentials and targeted
final thermal states may be relevant to recent research activity 
in the context of quantum engines \cite{Rezek,Abah,Deng}.

\end{document}